\documentclass[twocolumn,epjc3]{svjour3}
\smartqed  
\RequirePackage{graphicx}
\usepackage{amsmath,amssymb}
\usepackage{color}

\journalname{Eur. Phys. J. C}

\begin{document}

\title{\textbf{Anisotropic strange star with Tolman-Kuchowicz metric under $f(R,T)$ gravity}}
\author{Suparna Biswas\thanksref{e1,addr1} \and Dibyendu Shee\thanksref{e2,addr2}
\and B.K. Guha\thanksref{e3,addr1} \and Saibal Ray\thanksref{e4,{addr2},{addr3}}}

\thankstext{e1}{e-mail: sb.rs2016@physics.iiests.ac.in}
\thankstext{e2}{e-mail: dibyendu\_shee@yahoo.com}
\thankstext{e3}{e-mail: bkguhaphys@gmail.com}
\thankstext{e4}{e-mail: saibal@associates.iucaa.in}

\institute{Department of Physics, Indian Institute of Engineering Science and Technology, Shibpur, B. Garden, Howrah 711103, West Bengal, India\label{addr1}
\and Department of Physics, Government College of Engineering and Ceramic Technology, Kolkata 700010, West Bengal, India\label{addr2} \and Department of Natural
Sciences, Maulana Abul Kalam Azad University of Technology, Haringhata 741249, West Bengal, India\label{addr3}}

\date{Received: date / Accepted: date}

\maketitle

\begin{abstract}
In the current article, we study anisotropic spherically symmetric strange star under the background of $f(R,T)$ gravity
using the metric potentials of Tolman-Kuchowicz type~\cite{Tolman1939,Kuchowicz1968} as $\lambda(r)=\ln(1+ar^2+br^4)$ and $\nu(r)=Br^2+2\ln C$ which are free from singularity, satisfy stability criteria and also well behaved. We calculate the value of constants $a$, $b$, $B$ and $C$ using matching conditions and the observed values of the masses and radii of known samples.  To describe the strange quark matter (SQM) distribution, here we have used the phenomenological MIT bag model equation of state (EOS) where the density profile  ($\rho$) is related to the radial pressure ($p_r$) as $p_r(r)=\frac{1}{3}(\rho-4B_g)$.  Here quark pressure is responsible for generation of bag constant $B_g$. Motivation behind this study lies in finding out a non-singular physically acceptable solution having various properties of strange stars. The model shows consistency with various energy conditions, TOV equation, Herrera's cracking condition and also with Harrison-Zel$'$dovich-Novikov's static stability criteria. Numerical values of EOS parameter and the adiabatic index also enhance the acceptability of our model.
\end{abstract}

\section{Introduction}\label{sec:intro}
Einstein's General Relativity (GR) consists of conceptual ingenuity and mathematical elegance in its every steps. To describe a gravitating system and non-inertial frames from a classical view point at large scale regimes,  GR is a highly useful tool~\cite{Einstein1915}.
But inspite of its beauty, singularity makes it stagnant~\cite{Wheeler1962} in few cases. Also the theory is unable to behave rationally for the description of observational accelerating phase of the Universe. It cannot describe properly the cosmic dynamics of the Universe without considering the exotic form of matter-energy which is supposed to be known as dark matter and dark energy~\cite{Kamenshchik2001,Padmanabhan2002,Bento2002,Caldwell2002,Nojiri2003a,Nojiri2003b,Riess2004,Eisentein2005,Astier2006,Spergel2007}.
Einstein's GR does not consider the quantum nature of matter and cannot be quantized in a conventional way of renormalization. In this line of thinking it was shown~\cite{Utiyama1962} that inclusion of higher order curvature terms in Einstein-Hilbert action will make the system renormalizable in one loop. On the other hand, to include quantum corrections one has to consider higher order curvature invariants in the effective gravitational action at low energy regime~\cite{Birrell1982,Buchbinder1992,Vilkovisky1992}.

To overcome the above mentioned problems of GR, several modified or alternative theories of gravitation have been proposed from time to time by researchers.
The modification has been done by changing the geometrical part of the Einstein field equations (EFE), i.e., the action is modified by considering a generalized
functional form of gravitational Lagrangian density. Some of the well known relevant alternative theories of gravitation are $f(R)$
gravity~\cite{Nojiri2003c,Carroll2004,Allemandi2005,Nojiri2007,Bertolami2007},
$f(\mathbb{T})$ gravity~\cite{Bengochea2009,Linder2010}, $f(R,T)$
gravity~\cite{Harko2008,Bisabr2012,Jamil2012,Alvarenga2013,Shabani2013,Shabani2014,Zaregonbadi2016a,Shabani2017a,Shabani2017b},
$f(\mathbb{T},T)$ gravity~\cite{Momeni2014,Junior2015,Nassur2015,Salako2015,Gomez2016,Pace2017},
$f(G)$ gravity~\cite{Bamba2010a,Bamba2010b,Rodrigues2014} and $f(R,G)$ gravity~\cite{Nojiri2005} where $R$, $\mathbb{T}$, $T$ and
$G$ are scalar curvature, torsion scalar, trace of the energy momentum tensor and Gauss-Bonnet scalar respectively.

In the present work we are interested in $f(R,T)$ theory. This theory is developed by considering non-minimal coupling between the
Ricci Scalar $R$ and trace of the energy momentum tensor $T$. Harko et al.~\cite{Harko2011} introduced it first to tackle the problems in an efficient way. In this theory matter is considered
on an equal footing with geometry. As a result one can explore many interesting and novel features of the Universe, such as the role of dark
matter~\cite{Zaregonbadi2016b}. One can note that dependency of $T$ may come from the consideration of quantum effects or from the presence of an imperfect fluid.

Neutron stars have been the subject of immense study for the last few decades due to its small size (radius $\sim$ 11-15~km) and being the mass $\sim 1.4-2.0~M_\odot$ a tremendous
density ($\rho \sim 10^{17}$~kg/m$^3$), which distort the space-time geometry~\cite{Demorest2010}. They take birth from the gravitationally
collapsing massive star $(M > 8M_\odot)$, after type II supernovae explosion~\cite{Wilczek1999}. Neutron stars enriched with neutrons,
spin rapidly and often making several hundred rotations per second. Sometimes neutron stars form radio pulsars, emitting radio waves.

The tremendous pressure and density is probably responsible for phase transition of neutrons inside the neutron stars
to hyperon $(\Lambda, \Sigma, \Xi, \Delta, \Omega)$ and quark matter $(u, d, s)$. Cameron~\cite{Cameron1959} predicted that
conversion of nucleon into hyperon is energetically more favourable. Interior (specially core) of the neutron star contains quark matter since
they become free of interaction due to high energy density and extreme asymptotic momentum transfer. The idea that quark matter may exist in the
core of the neutron star, has been suggested by several scientists~\cite{Ivanenko1965,Itoh1970,Fritzsch1973,Baym1976,Keister1976,Chapline1977,Fechner1978}.

The energy level of the hyperon at the Fermi-surface becomes higher than its rest mass, due to the tremendous density and as a
result these particles could be deconfined into strange quarks. Though quark stars may contain up $(u)$, down $(d)$ and strange
$(s)$ quarks but mostly they contain strange $(s)$ quarks~\cite{Itoh1970,Bodmer1971,Witten1984,Haensel1986,Alcock1986} since strange quarks
are the most stable ones.
Under some conditions, up $(u)$ and down $(d)$ quarks can also be transformed into strange $(s)$ quarks. So once up $(u)$ and down $(d)$ quarks of the quark star convert
into strange matter, the entire quark matter get converted into strange matter. In this way neutron stars may be totally converted into strange quark star~\cite{Pagliara2013}.
The neutron stars may very well contain quark matter in their cores, which ought to be in a
color superconducting state~\cite{Rajagopal2001,Alford2001,Alford1998,Rapp1998}. But still now, no one has been able to predict
(theoretically or experimentally) the exact density at which expected phase transition to quark matter occurs.

Quantum Chromodynamics (QCD) deals with the quark confinement mechanism. At asymptotic density, the ground state of QCD
with a vanishing strange quark mass is the color-flavor-locked (CFL) phase. According to Chodos et al.~\cite{Chodos1974}, a
strongly interacting particle can be defined as a finite region of space, confined with fields and bag constant is such a candidate.
As a result, bag constant affects the energy momentum tensor of the star and even the spacetime geometry. According to the MIT bag model, the universal pressure $B_g$ known as bag constant, is responsible for quark confinement and it is defined as the difference between energy density of the perturbative and non-perturbative quantum chromodynamics vacuum. According to Farhi~\cite{Farhi1984}
and Alcock~\cite{Alcock1986}, for a stable strange quark matter (SQM), the bag constant should be within the following
range (55-75)~MeV/fm$^3$ under GR. However CERN-SPS and RHIC~\cite{Burgio2002} provided a wide range of this constant.

Abbas et al.~\cite{Abbas2015} using MIT bag model, investigated anisotropic charged strange stars in $f(T)$ gravity employing the diagonal tetrad field of static space-time. Biswas et al.~\cite{Biswas2018}, have established a new model for highly compact anisotropic strange star using the metric potentials, given by Krori-Barua~\cite{Krori1975} under $f(R,T)$ gravity. Besides these, several studies are available in literature survey
~\cite{Madsen1999,Rapp2000,Madsen2002,Paulucci2014,Isayev2015,Arbanil2016,Lugones2017} based on strange star.

Inhomogeneous matter distribution and its evolution changes the fabric of the spacetime of compact objects and leads to the anisotropic features. Anisotropy plays an important role to study the stellar properties. Anisotropic pressure measures the difference between radial component of pressure and tangential component of pressure. Ruderman~\cite{Ruderman1972} first argued that at very high density $(> 10^{15}$~gm/cc) anisotropy in pressure arises and we have to treat nuclear interactions relativistically. Different physical properties of a stellar object like energy density, total mass, gravitational redshift and frequency of oscillation of the fundamental mode are affected by the anisotropy. The reasons behind the anisotropy are provided by different scientists in their research work from time to time. Usov~\cite{Usov2004} suggested that strong electric field is the producer of anisotropy. According to Weber~\cite{Weber1999}  the
anisotropy in pressure inside a strange star is due to immense magnetic field of the neutron star. Sokolov~\cite{Sokolov1980} argued that  different types of phase transitions can be taken as responsible for this anisotropy. Also pion condensation can be taken as one of the reason behind the anisotropy, predicted by Sawyer~\cite{Sawyer1972}. According to Kippenhahn and Weigert~\cite{Kippenhahn1990}, the anisotropy may arise due to presence of a type 3A superfluid or existence of solid stellar core. The literature survey regarding anisotropic fluid sphere are provided in the following references~\cite{Bower1974,Bayin1982,Krori1984,Maharaj1989,Ivanov2002,Schunck2003,Mak2003,Varela2010,Rahaman2010,Rahaman2011,Rahaman2012,Maurya2012,Kalam2012,Maurya2015,Malaver2015,Pant2015,Maurya2016,Shee2016,Shee2017}.

In this background, we shall summarize our research work as follows. Basic mathematics behind $f(R,T)$ gravity has been discussed in Sect.~\ref{sec 2}. Solutions to EFEs for strange star and the proper choice of EOS for SQM distribution have been explained in Sect.~\ref{sec 3}. Sect.~\ref{sec 4} contains the boundary conditions from whics we have calculated all the model parameters. In Sect.~\ref{sec5}, different physical features like density and pressure (\ref{subsec5.1}), energy conditions (\ref{subsec5.2}), have studied along with different stability criteria like Herrera's cracking conditions~(\ref{subsubsec5.3.1}), TOV equation~(\ref{subsubsec5.3.2}), adiabatic index~(\ref{subsubsec5.3.3}), equations of state parameter~(\ref{subsubsec5.3.4}) and Harrison-Zel$'$dovich-Novikov static
stability criteria~(\ref{subsubsec5.3.5}) etc. The effective mass, compactness, i.e., the mass-radius ratio from which one can calculate the gravitational as well as the surface redshift, has been discussed in Sect.~\ref{sec6}. Finally, in Sect.~\ref{sec7}, we have provided an overall conclusion of our detailed study from different aspect of strange star model.

\section{Mathematics behind $f(R,T)$ gravity}\label{sec 2}
Following Harko~\cite{Harko2011}, we can represent action in $f(R,T)$ theory as
\begin{equation}
S=\int d^4x \pounds_m\sqrt{-g} +\frac{1}{16\pi}\int d^4x f(R,T)\sqrt{-g}.\label{eq3.1}
\end{equation}

Here the arbitrary function $f(R,T)$ contains the Ricci scalar $R$ as well as $T$, the trace of the energy momentum
tensor, $\pounds_m$ is the matter Lagrangian density which represents the possibility
of non-minimal coupling between matter and geometry, $g$ is the determinant of $g_{\mu\nu}$ metric ($c = G = 1$).

Varying the action (\ref{eq3.1}) w.r.t. $g_{\mu\nu}$ metric, we can derive modified EFEs for $f(R,T)$ gravity as
\begin{eqnarray}
&\qquad\hspace{-1cm}f_R(R,T)R_{\mu\nu}-\frac{1}{2}g_{\mu\nu}f(R,T)+f_R(R,T)(g_{\mu\nu}\square-\nabla_\mu\nabla_\nu) \notag \\
&\qquad\hspace{-1cm}=8\pi\mathcal{T}_{\mu\nu}-\mathcal{T}_{\mu\nu}f_T(R,T)-\Theta_{\mu\nu}f_T(R,T),\label{eq3.2}
\end{eqnarray}
where $f_T(R,T)=\frac{\partial f(R,T)}{\partial T}$ , $f_R(R,T)=\frac{\partial f(R,T)}{\partial R}$,
$\square\equiv\frac{\partial_\mu(\sqrt{-g}g^{\mu\nu}\partial_\nu)}{\sqrt{-g}}$, $R_{\mu\nu}$
represents the Ricci tensor, $\nabla_\mu$ is the covariant derivative w.r.t. the symmetry connected to $g_{\mu\nu}$,
$\Theta_{\mu\nu}=g^{\alpha{\beta}}\delta \mathcal{T}_{\alpha{\beta}}/\delta g^{\mu\nu}$ and
$\mathcal{T}_{\mu\nu}=g_{\mu\nu}\pounds_m-2\frac{\partial\pounds_m}{\partial g^{\mu\nu}}$ represents the stress-energy tensor.

From Eq. (\ref{eq3.2}), the covariant divergence becomes~\cite{Barrientos2014}
\begin{eqnarray}
&\qquad\hspace{-0.5cm}\nabla^\mu \mathcal{T}_{\mu\nu}=\frac{f_T(R,T)}{8\pi-f_T(R,T)}[(\Theta_{\mu\nu}+\mathcal{T}_{\mu\nu})\nabla^\mu\ln f_T(R,T) \notag \\
&\qquad\hspace{-0.8cm}-\frac{1}{2}g_{\mu\nu}\nabla^\mu T+\nabla^\mu\Theta_{\mu\nu}].\label{eq3.3}
\end{eqnarray}

Eq. (\ref{eq3.3}) states that in $f(R,T)$ theory of gravity, energy-momentum tensor is not
conserved where as it remains conserved in general relativity.

For a perfect anisotropic fluid, the energy-momentum tensor takes the following form
\begin{equation}
\mathcal{T}_{\mu\nu}=(\rho+p_t)u_\mu u_\nu-p_t g_{\mu\nu}+(p_r-p_t)v_\mu v_\nu, \label{eq3.4}
\end{equation}
with $u^\mu\nabla_\nu u_\mu=0$ and $u^{\mu}u_{\mu}= 1$. Here $\rho(r)$,
$p_r(r)$, $p_t(r)$, $u_\mu$, $v_\mu$ stand for the energy density, radial pressure and
tangential pressure, four-velocity and radial four-vector respectively for a static fluid source. Besides
these, we have another condition $\Theta_{\mu\nu}=-2\mathcal{T}_{\mu\nu}-pg_{\mu\nu}$.

Following Harko et al.~\cite{Harko2011}, $f(R,T)$ takes the form
\begin{equation}
f(R,T)=R+2\chi T.  \label{eq3.5}
\end{equation}

Here $\chi$ is a constant arises due to the coupling between matter and geometry in modified gravity.

Substituting Eq.~(\ref{eq3.5}) in Eq.~(\ref{eq3.2}), we get
\begin{equation}\label{eq3.6}
G_{\mu\nu}=8\pi \mathcal{T}_{\mu\nu}+\chi T g_{\mu\nu}+2\chi(\mathcal{T}_{\mu\nu}+pg_{\mu\nu}).
\end{equation}
where $G_{\mu\nu}$ is known as the Einstein tensor. Now, substituting $\chi=0$ in the Eq.~(\ref{eq3.6}),
we get different results which represent the results of GR.

Combination of Eqs. (\ref{eq3.3}) and (\ref{eq3.5}) gives the following result
\begin{equation}
(8\pi+2\chi)\nabla^{\mu}\mathcal{T}_{\mu\nu}=-2\chi\left[\nabla^{\mu}(pg_{\mu\nu})+\frac{1}{2}g_{\mu\nu}\nabla^{\mu}T\right].\label{eq3.7}
\end{equation}

Incorporation of $\chi=0$ in Eq. (\ref{eq3.7}) shows the invariance of energy-momentum tensor in general relativity.

\section{Solution of Einstein's field equations} \label{sec 3}
To study a strange star, we consider the static, spherically symmetric stellar configuration, for which
the line element of the interior space-time, can be described as the following metric:
\begin{equation}
ds^2 =e^{\nu(r)}dt^2-e^{\lambda(r)}dr^2-r^2(d\theta^2+sin^2\theta d\phi^2).\label{eq3.8}
\end{equation}

Here the metric potentials $\lambda(r)$ and $\nu(r)$ are chosen as $\lambda(r)=\ln(1+ar^2+br^4)$ and $\nu(r)=Br^2+2\ln C$,
known as Tolman-Kuchowicz metric potentials~\cite{Tolman1939,Kuchowicz1968}. The constants $a$, $b$, $B$ and $C$
can be determined using matching conditions. The non-zero components of energy-momentum tensor can be written as
$\mathcal{T}_0^0=\rho(r)$, $\mathcal{T}^1_1=-p_r(r)$ and $\mathcal{T}^2_2=\mathcal{T}^3_3=-p_t(r)$.

For an anisotropic spherically symmetric uncharged stellar system, combining Eqs.~(\ref{eq3.4}),~(\ref{eq3.6}) and~(\ref{eq3.8}),
we get the EFE as follows
\begin{eqnarray}
&\qquad\hspace{-1.2cm}\frac{e^{-\lambda}}{r^2}(-1+e^\lambda+r\lambda')=8\pi\rho+2\chi\left[2\rho-\frac{p_r+2p_t}{3}\right] =8\pi\rho^{eff},~~\label{eq3.9} \\
&\qquad\hspace{-1.2cm}\frac{e^{-\lambda}}{r^2}(1-e^{\lambda}+r\nu')=8\pi p_r-2\chi\left[\rho-\frac{4p_r+2p_t}{3}\right]=8\pi p_r^{eff},~~~\label{eq3.10} \\
&\qquad\hspace{-2.8cm}\frac{e^{-\lambda}}{4r}\left[2(\nu'-\lambda')+(2\nu''+\nu'^2-\lambda'\nu')r\right]= \notag \\
&\qquad\hspace{-1cm}8\pi p_t-2\chi\left[\rho-\frac{p_r}{3}-\frac{5p_t}{3}\right]=8\pi p_t^{eff}.\label{eq3.11}
\end{eqnarray}
where $\prime$ expresses the differentiation of the respective parameters w.r.t. $r$.

The strange quark matter distribution inside the strange star has been generated by the simplest phenomenological MIT bag model
EOS~\cite{Chodos1974,Farhi1984,Alcock1986}. The quark pressure of SQM, considering the quarks are non-interacting, massless and
including all the corrections of energy and pressure, can be defined as
\begin{equation}
p_r(r)=\sum_{f=u,d,s} p^{f}-B_{g}, \label{eq3.12}
\end{equation}
where $p^{f}$ is the individual pressure of all three flavors of quarks and $B_{g}$ is the bag Constant.

The relation between the individual quark pressure $p^{f}$ and energy density of individual quark flavor is given by $p^{f}=\frac{1}{3}\rho^{f}$. Therefore deconfined quarks inside the bag have the total energy $\rho$ as follows
\begin{equation}
\rho=\sum_{f=u,d,s} \rho^{f}+B_g. \label{eq3.13}
\end{equation}

Using Eqs.~(\ref{eq3.12}) and~(\ref{eq3.13}) we have the MIT bag EOS as
\begin{equation}
p_r(r)=\alpha\left[\rho(r)-4B_g\right], \label{eq3.14}
\end{equation}
where $\alpha$ is a constant having numerical value 0.28 for the massive strange quarks with mass $250$~MeV
and $\frac{1}{3}$ for massless strange quarks.

MIT bag model equation of state (EOS) for strange quark matter (SQM) can be written as
\begin{equation}
p_r(r)=\frac{1}{3}\left[\rho(r) -4{B_g}\right]. \label{eq3.12}
\end{equation}

Inserting $\lambda(r)=\ln(1+ar^2+br^4)$ and $\nu(r)=Br^2+2\ln C$ in Eqs. (\ref{eq3.9})-(\ref{eq3.12}), we get
\begin{eqnarray}
&\qquad\hspace{-2cm}\rho_r^{eff}=\frac{1}{X}\left[12b^2{B_g}r^8Y+24abB_gr^6Y-4B^2b\chi{r}^6+ \right. \nonumber \\
&\qquad\hspace{-1.3cm}\left.12a^2B_gr^4Y-4B^2a\chi r^4+24b{B_g}r^4Y+36B\pi br^4+23Bb\chi r^4\right.\nonumber \\
&\qquad\hspace{-2cm}\left.+24a{B_g}r^2Y-4B^2\chi r^2+36B\pi ar^2+19Ba\chi{r}^2+72\pi br^2\right.\nonumber \\
&\qquad\hspace{-2.2cm}\left.+54b\chi{r}^2+12{B_g}Y+36B\pi+15B\chi+36\pi a+27a\chi\right], \label{eq3.13}\\
&\qquad\hspace{-2.2cm}p_r^{eff}=\frac{1}{X}\left[-12b^2{B_g}r^8Y-24ab{B_g}{r}^6Y+4B^2b\chi{r}^6-\right.\nonumber \\
&\qquad\hspace{-1.4cm}\left.12a^2B_gr^4Y+4B^2a\chi{r}^4-24br^4B_gY+12B\pi b{r}^4-3Bb\chi{r}^4  \right.\nonumber\\
&\qquad\hspace{-2.5cm}\left.-24ar^2B_gY+Br^2(4B\chi+12a\pi+a\chi)+24\pi br^2\right.\nonumber\\
&\qquad\hspace{-2.5cm}\left.-14b\chi{r}^2-12B_gY+12\pi(B+a)+5B\chi-7a\chi\right] , \label{eq3.14}\\
&\qquad\hspace{-3cm}p_t^{eff}=\frac{B^2br^6+B^2ar^4+B^2r^2+Bar^2-2br^2+2B-a}{8\pi(br^4+ar^2+1)^2}.\label{eq3.15}
\end{eqnarray}
where, $X=16\pi(12\pi+5\chi)(br^4+ar^2+1)^2$, $Y=(16\pi^2+16\pi\chi+3\chi^2)$

The anisotropic stress can be obtained as
\begin{eqnarray}
&\qquad\hspace{-1.2cm}\Delta=[p_t^{eff}- p_r^{eff}]=\frac{3}{X}\left[4b^2{B_g}r^8Y+8ab{B_g}r^6Y+8B^2\pi br^6\right.\nonumber\\
&\qquad\hspace{-1.1cm}\left.+2B^2b\chi{r}^6+4a^2{B_g}{r}^4Y+8B^2\pi ar^4+2B^2a\chi{r}^4+8b{B_g}{r}^4Y\right.\nonumber\\
&\qquad\hspace{-1.3cm}\left.-4B\pi br^4+Bb\chi r^4+8a{B_g}{r}^2Y+8B^2\pi r^2+2B^2\chi{r}^2\right.\nonumber\\
&\qquad\hspace{-1.4cm}\left.+4B\pi ar^2+3Ba\chi{r}^2-24\pi br^2-2b\chi r^2+4{B_g}Y+12B\pi\right.\nonumber\\
&\qquad\hspace{-.2cm}\left.+5B\chi-12\pi a-a\chi\right], \label{eq3.16}
\end{eqnarray}
where, $X=16\pi(12\pi+5\chi)(br^4+ar^2+1)^2$, $Y=(16\pi^2+16\pi\chi+3\chi^2)$.

\section{Matching conditions}\label{sec 4}

\subsection{Interior space-time}\label{subsec4.1}
The effective density at the center from Eq.~(\ref{eq3.13}) is
\begin{eqnarray}
&\qquad\hspace{-4.5cm}\rho_0^{eff} =\rho^{eff}(r=0)  .\nonumber\\
&\qquad\hspace{-.7cm}=\frac{192\pi^2{B_g}+192\pi{B_g}\chi+36{B_g}\chi^2+36B\pi+15B\chi+36\pi a+27a\chi}{16\pi(12\pi+5\chi)}.\label{eq3.17}
\end{eqnarray}

Here we assume the coupling constant $\chi=1$. Anisotropic condition, i.e., at the center ($r=0$) radial pressure balances the tangential
pressure as $p_r^{eff}(r=0)=p_t^{eff}(r=0)$. Using these conditions, we get
\begin{eqnarray}
&\qquad\hspace{-.5cm}{B_g}=\frac{12(a-B)\pi-5B+a}{64\pi^2+64\pi+12}.\label{eq3.18}
\end{eqnarray}

\subsection{Exterior space-time}\label{subsec4.2}
As there is no mass in exterior region, matter geometry coupling constant $\chi$ becomes zero, leading all the components of the energy momentum tensor $\mathcal{T}_{\mu\nu}=(\rho+p_t)u_\mu u_\nu-p_t g_{\mu\nu}+(p_r-p_t)v_\mu v_\nu$ to be zero. So at the exterior Schwarzschild solution is as follows
\begin{eqnarray}
&\qquad\hspace{-.9cm}ds^2=\left(1-\frac{2M}{r}\right)dt^2-\frac{dr^2}{\left(1-\frac{2M}{r}\right)}-r^2(d\theta^2+\sin^2\theta d\phi^2),\label{eq3.19}
\end{eqnarray}
where $M$ represents total mass of the stellar system. From matching conditions, the metric coefficients $g_{tt}$, $g_{rr}$ and
$\frac{\partial g_{tt}}{\partial r}$ are continuous between the interior and exterior regions, at the surface, i.e., $r=\Re$
(where $\Re$ is the radius). Hence, by comparing Eqs. (\ref{eq3.8}) and (\ref{eq3.19}), we get
\begin{eqnarray}\label{eq3.20}
\begin{aligned}
&g_{tt}=1-\frac{2M}{\Re}=e^{B\Re^2+2\ln C}, \\
&g_{rr}=1-\frac{2M}{\Re}=e^{(1+ar^2+br^4)},  \\
&\frac{\partial g_{tt}}{\partial r}=\frac{2M}{\Re^2}=2B\Re e^{B\Re^2+2\ln C}.
\end{aligned}
\end{eqnarray}

Again, at the surface $(r=\Re)$ effective radial pressure vanishes, i.e., $p_r^{eff}(r=\Re)=0$. Comparing this condition
with Eq.~(\ref{eq3.18}) and using Eq.~(\ref{eq3.20}) we can represents the unknown constants $a, b, B, C$ and ${B_g}$
in terms of $M$ and $\Re$.

\begin{figure}
\centering
\includegraphics[width=6cm]{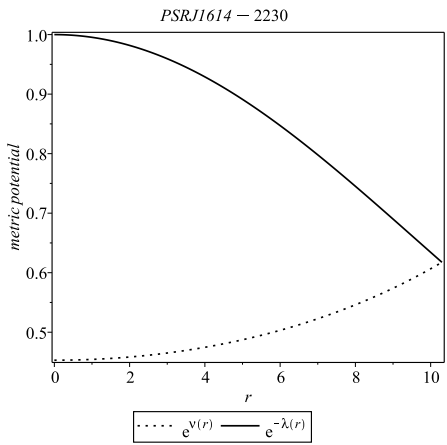}
\caption{Variation of $e^{-\lambda}$, $e^{\nu}$ w.r.t. the radial coordinate $r$ for the strange star candidate $PSRJ~1614-2230$.}\label{3.pot}
\end{figure}

\begin{figure}
\centering
\includegraphics[width=6cm]{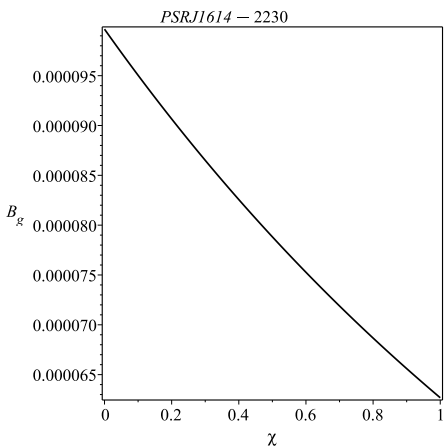}
\caption{Variation of $B_g$ w.r.t $\chi$ for the strange star candidate $PSRJ~1614-2230$.}\label{3.Bg}
\end{figure}

Variations of the metric potentials $e^{\nu}$ and $e^{\lambda}$ with the radial coordinate (vide Fig.~\ref{3.pot}) imply that
$e^{\nu(r)}|_{r=0}$ has a non-zero positive value and ${e^{\lambda(r)}|_{r=0}}=1$. These are the necessary conditions
for the solution to be free from physical and geometrical singularity. The metric potentials increase non-linearly from the
center of the star and achieve maximum value at the surface of the star. Variation of $B_g$ with the coupling constant $\chi$
has also shown in Fig.~\ref{3.Bg} (right panel). From the display, one can note that $\chi$ effectively reduces the $B_g$ value. Here,
for simplification, we have assumed $\chi=1$ for which $B_g \sim$ 47~MeV/fm$^3$. But, while approaching $\chi\rightarrow0$, $B_g$ increases and reaches it's maximum at $\chi=0$ as $\sim$ 68~MeV/{fm}$^3$.

\section{Physical characteristics of the proposed model}\label{sec5}
In this section we shall check the physical validity of our model and also whether our proposed fluid distribution is in stable equilibrium or not.

\subsection{Density and pressure}\label{subsec5.1}
\begin{figure}
\centering
\includegraphics[width=6cm]{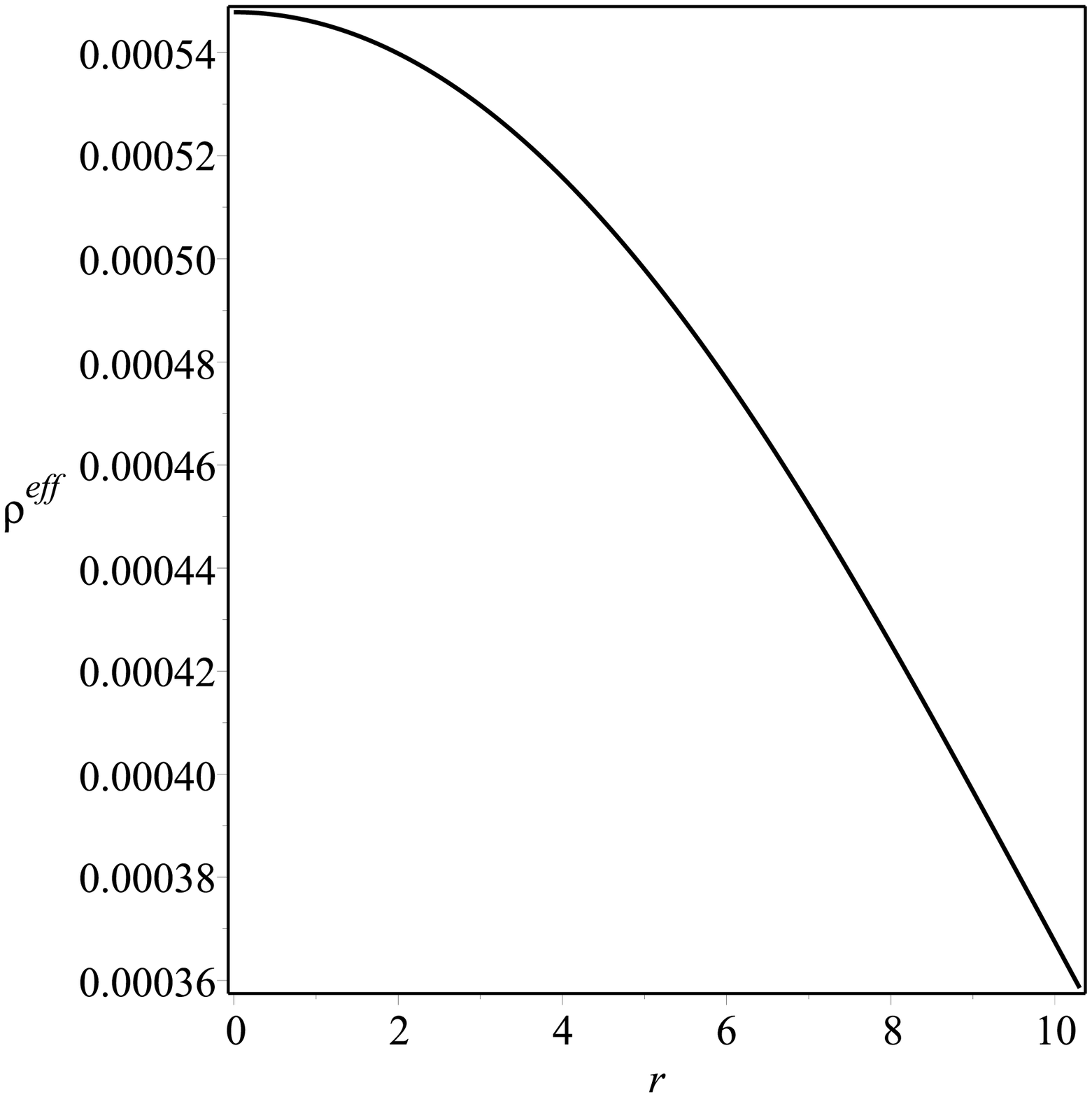}
\includegraphics[width=6cm]{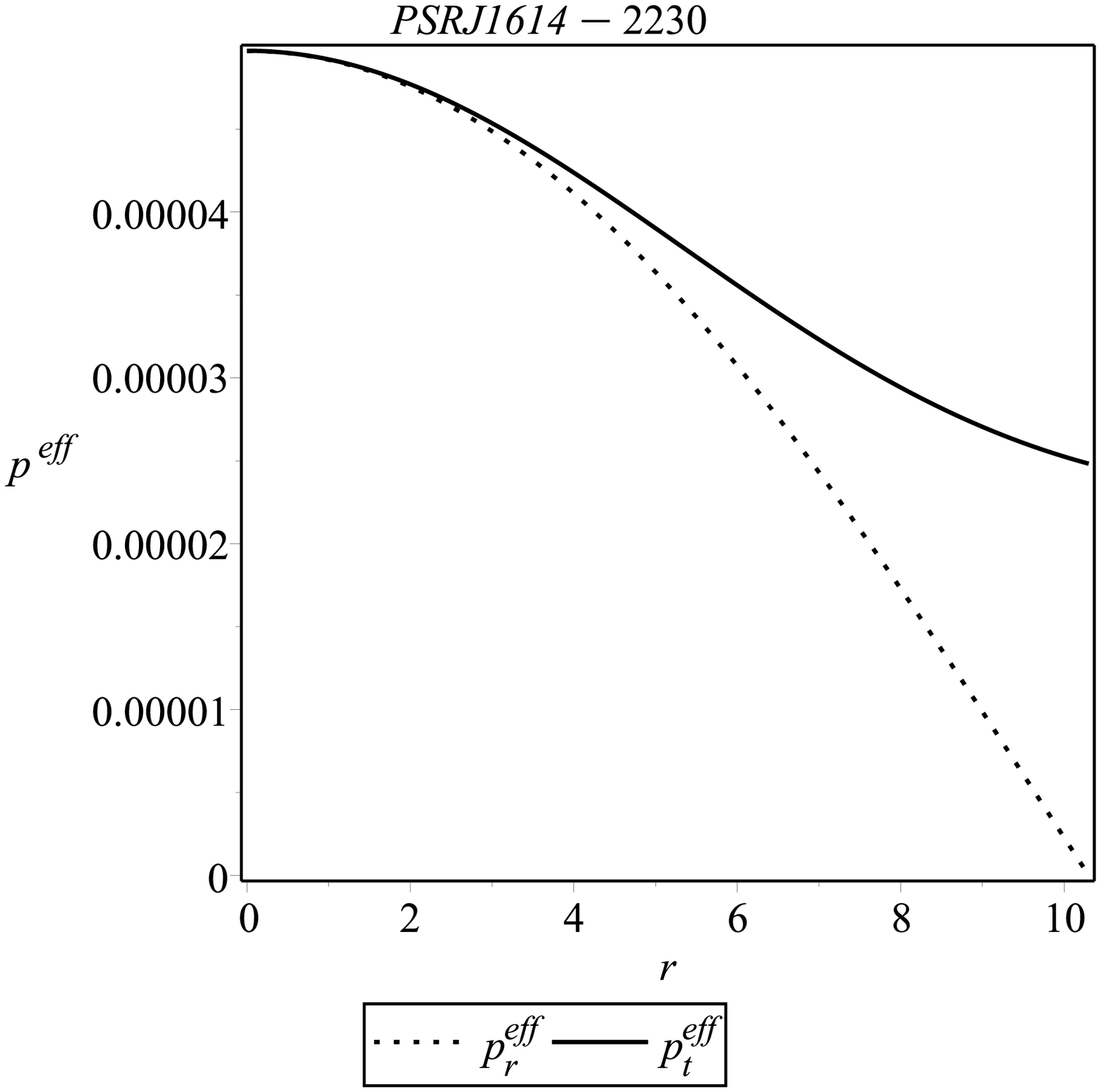}
\includegraphics[width=6cm]{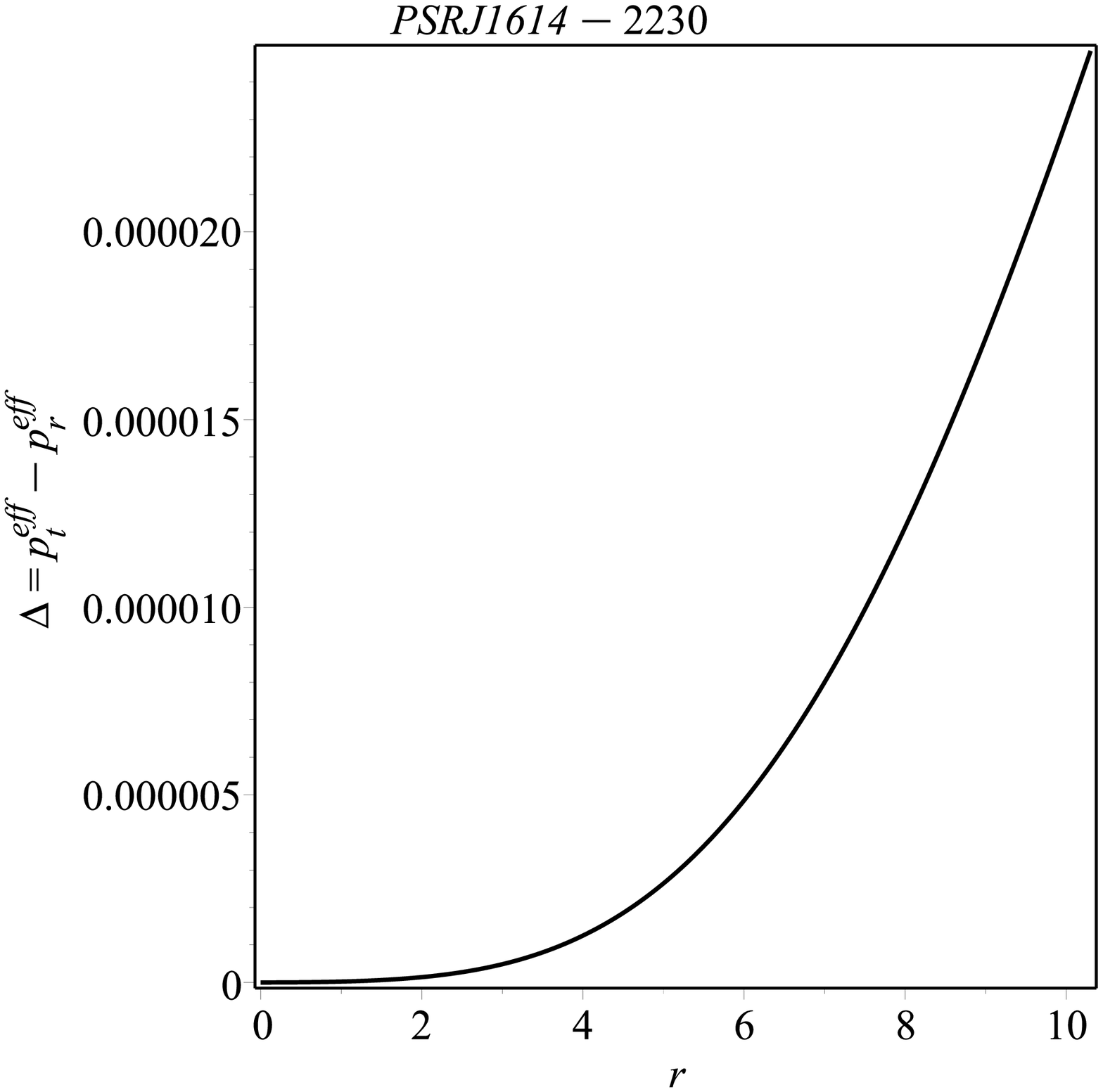}
\caption{Variation of $\rho^{eff}$ (upper panel), $p_r^{eff}$, $p_t^{eff}$ (middle panel) and
anisotropic stress (lower panel) w.r.t. the radial coordinate $r$ for the strange star candidate $PSRJ$~1614-2230.}\label{3.pressue}
\end{figure}

We know that the density and pressure are the most important physical quantity in the study of a stellar system. Since we consider
anisotropic fluid distribution in the modified gravity in the specific form $f(R,T)$, so we have to consider the effective value of those quantities.
Eqs.~(\ref{eq3.13})-(\ref{eq3.15}) provide us the effective density, effective radial pressure and effective tangential pressure. Since, as an example for strange star candidate, we have considered $PSRJ~1614-2230$, it will be most effective if we look at the numerical values of those effective quantities for this star. From Table~\ref{Table2} it is noticeable that effective central density is $7.412 \times 10^{14}$ {gm/cm}$^3$ where as its value at the surface of the star is $4.85\times10^{14}$ {gm/cm}$^3$. Though the decreasing nature of the effective density is very prominent but very high matter density throughout the stellar system make it as a possible candidate of ultra dense strange quark star~\cite{Glendenning1997,Herzog2011}. The graphical variation of $\rho^{eff}$ (Fig~\ref{3.pressue},~left panel) demonstrates that it is maximum at the center and decreases gradually with the radius of the star.

The variation of the effective radial pressure $p_r^{eff}$ as well as the transverse pressure $p_t^{eff}$ (Fig~\ref{3.pressue},
~middle panel) also maintain the same pattern, i.e., they are maximum at the center and decreases nonlinearly with the radius
of the star. Their finite value at the center of the star makes it a singularity free model of strange star. The graphical
variation of the pressures clearly indicate that at the surface, effective radial pressure vanishes prominently whereas
effective tangential pressure retain a finite value which make the shape of the star spheroidal in nature~\cite{Quevedo1989,Chifu2012,Shee2017}. The origin of this spheroidicity makes it a topic of future research.

The anisotropic stress is represented in Eq.~(\ref{eq3.16}). The beauty of this expression is that the anisotropy will be directed inward when $\Delta<0$, i.e. $p_t^{eff}<p_r^{eff}$ whereas $\Delta>0$, i.e. $p_t^{eff}>p_r^{eff}$ implies that direction of the anisotropy will be outward~\cite{Hossein2012}. From the graphical
 variation of the anisotropic stress (Fig~\ref{3.pressue},~right panel), it is clear that at the origin anisotropy vanishes,
 i.e. at the center the effective perpendicular pressures are equal. Also we can observe that anisotropy increases nonlinearly with the
 increase of the radius of the star. The positive value of the anisotropic stress helps to construct more compact object as described by
 Gokhroo and Mehra~\cite{Gokhroo1994}. The graphical representation of anisotropic stress shows that it achieves maximum value at the
 surface of the star, which according to the Deb et al.~\cite{Deb2017} is an inherent property of any ultra dense compact object.

\subsection{Energy Conditions}\label{subsec5.2}
The distribution of matter in the space-time, as measured by an observer, is termed as energy conditions. The positivity
of these conditions implies the flow of matter should be along null or time-like world line. Our proposed anisotropic fluid
distribution will be consistent with the energy conditions, i.e., the null energy condition (NEC),
weak energy condition (WEC), strong energy condition (SEC) and dominant energy condition (DEC) only if it satisfies the
following inequalities simultaneously:
\begin{eqnarray} \label{eq3.26}
\begin{aligned}
NEC &:\rho^{eff} \geq 0,  \\
WEC &:\rho^{eff}+p_r^{eff} \geq 0 , \rho^{eff}+p_t^{eff} \geq 0, \\
SEC &:\rho^{eff}+p_r^{eff}+2p_t^{eff} \geq 0,         \\
DEC &:\rho^{eff}-|p_r^{eff}| \geq 0 , \rho^{eff}-|p_t^{eff}| \geq 0.
\end{aligned}
\end{eqnarray}

\begin{figure}
\centering
\includegraphics[width=6cm]{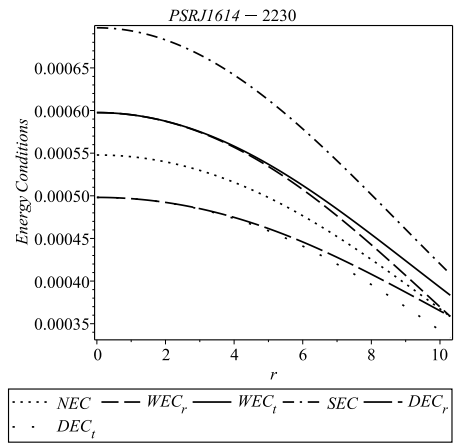}
\caption{Variation of the different energy conditions w.r.t. the radial coordinate $r$ for the strange star candidate $PSRJ$~1614-2230.}\label{3.energy}
\end{figure}

But several matter configurations violate the strong energy condition, e.g. for a scalar field with a positive potential
and any cosmological inflationary process~\cite{Hawking1973} can violate this condition. In such a situation we have to use an alternative
theory of gravity since violation of SEC will violate the classical regime of GTR. From the graphical variation (Fig.~\ref{3.energy}) it is
clear that our model satisfies all the energy conditions for coupling constant $\chi=1$ which will represent a stable anisotropic strange star configuration of our consideration.

\subsection{Stability Criteria}\label{subsec5.3}

\subsubsection{Herrera's concept for stability analysis}\label{subsubsec5.3.1}
The stability of our model can be checked by the help of causality condition of Herrera. According to this condition, square of the
radial and tangential speed of sound for anisotropic fluid distribution should follow the results $0<v^2_{rs}<1$ and $0<v^2_{ts}<1$.
Another concept for checking the stability of a stellar system is known as cracking concept. It appears when the equilibrium configuration of
the fluid distribution has been perturbed. According to cracking concept the region for which square of the transverse speed of sound is smaller than the square of the radial speed of sound, is a potentially stable region~\cite{Herrera1979,Herrera1992,Chan1993,Abreu2007,Andreasson2009}.
For the stable matter distribution Herrera~\cite{Herrera1992} and Andr{\'e}asson~\cite{Andreasson2009} demands that $\vert v^2_{rs}-v^2_{ts} \vert \leq 1$. The cracking has to appear either from the anisotropy of fluid distribution or due to the emission of incoherent radiation.

\begin{figure}
\centering
\includegraphics[width=6cm]{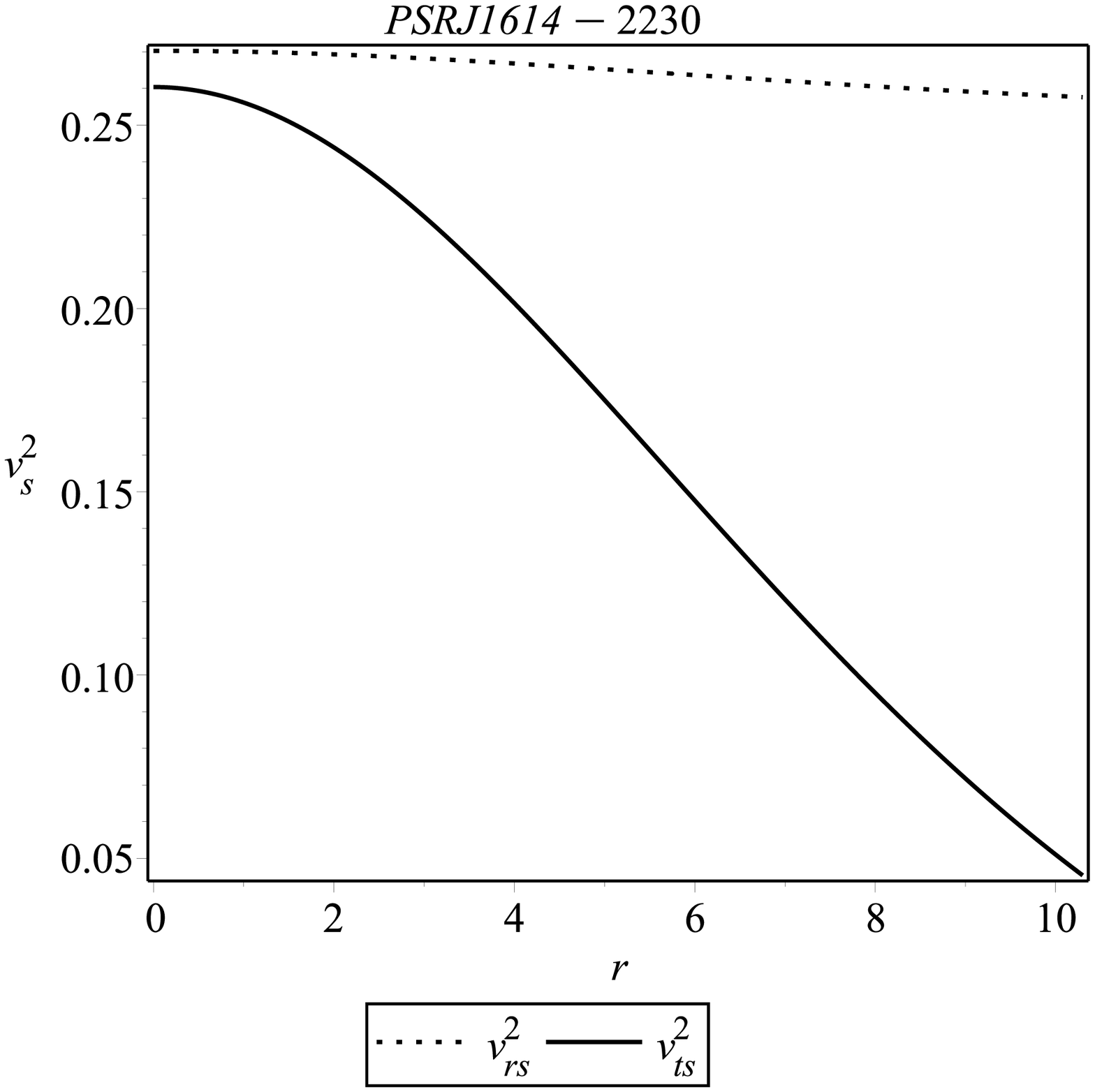}
\includegraphics[width=6cm]{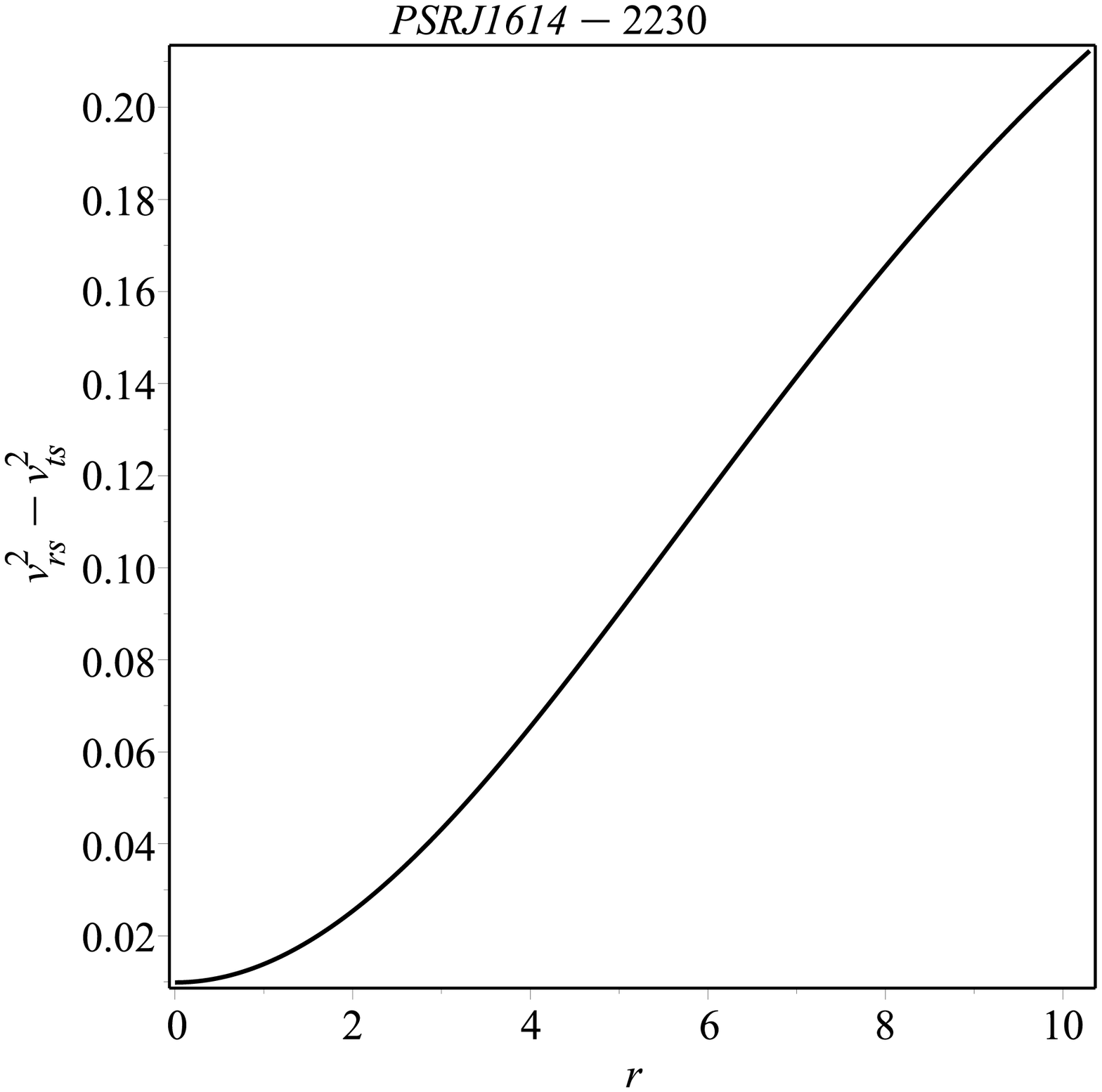}
\caption{Variation of $v^2_{rs}$ and $v^2_{ts}$ (upper panel), $\vert v^2_{rs}-v^2_{ts}\vert$ (lower panel) w.r.t.
the radial coordinate $r$ for the strange star candidate $PSRJ$~1614-2230.}\label{3.Hererra}
\end{figure}

In Fig.~\ref{3.Hererra}, variations of $v^2_{rs}$ and $v^2_{ts}$ with respect to the radius of the star has been shown and it
is clear that they remain within their specified range $(0,1)$ throughout the stellar system, which confirm the causality
condition. Also it is observed clearly that the term $\vert v^2_{rs}-v^2_{ts}\vert$ confirm the cracking concept.

\subsubsection{The generalized Tolman-Oppenheimer-Volkoff equation}\label{subsubsec5.3.2}
The modified form of the energy momentum tensor in the framework of $f(R,T)$ gravity is presented in Eq.~(\ref{eq3.7}).
Hence the modified form of the Tolman-Oppenheimer-Volkoff equation, which describes the stability of the proposed model, is given by
\begin{eqnarray}
&\qquad\hspace{-3cm}-\frac{(\rho+p_r)\nu'}{2}-\frac{dp_r}{dr}+\frac{2}{r}(p_t-p_r) \nonumber \\
&\qquad\hspace{+1cm}+\frac{\chi}{(8\pi+2\chi)}\left[\frac{d}{dr}(\frac{p_r+2p_t}{3})-\frac{d\rho}{dr}\right]=0.\label{eq3.25}
\end{eqnarray}

Here the first term represents the gravitational force $(F_g)$, second term denotes the hydrodynamic force $(F_h)$ and third term indicates the anisotropic force $(F_a)$. The fourth term arises due to coupling between the matter and the geometry, which can be termed as the force due to modified gravity $(F_{mg})$. So we can conclude that sum of all the forces are zero, i.e., $F_g+F_h+F_a+F_{mg}=0$, which implies that our system is completely stable.

\begin{eqnarray}
&\qquad\hspace{-5.5cm}F_g=-\frac{(Bbr^4+Bar^2+2br^2+B+a)Br}{(br^4+ar^2+1)^2\xi}, \label{eq3.25a}\\
&\qquad\hspace{-1.6cm}F_h=\frac{r(2Bb^2r^6+3Babr^4+6b^2r^4+Ba^2r^2+2Bbr^2+6abr^2+Ba+2a^2-2b)}{2(br^4+ar^2+1)^3\xi},\label{eq3.25b} \\
&\qquad\hspace{-1.3cm}F_a=\frac{2}{r}\left[\frac{8b^2\chi{B_g}\xi r^8+12b\xi(\frac{16}{3}a\chi{B_g}+B^2)r^6+Zr^4+Yr^2+8{B_g}\chi^2+X\chi+P}{2(12\pi+5\chi)\xi(br^4+ar^2+1)^2}\right.\nonumber \\
&\qquad\hspace{-1.3cm}\left.-\frac{Bbr^4+Bar^2+2br^2+B+a}{4(br^4+ar^2+1) ^2\xi}+B_g\right],\label{eq3.25c} \\
&\qquad\hspace{-1.8cm}F_{mg}=-\frac{\left[B^2b^2r^8+bB(Ba-4b)r^6-3b(Ba+6b)r^4+Pr^2-B^2+Q\right]r\chi}{(12\pi+5\chi)\xi(br^4+ar^2+1)^3}\label{eq3.25d},
\end{eqnarray}
where, $\xi=(\chi+4\pi)$, $P=(-B^2a+(-a^2+4b)B-18ba)$, $Z=[8{B_g}(a^2+2b)\chi^2+\left(32{B_g}(a^2+2b)\pi+3B^2a+4bB\right)\chi+12B^2a\pi]$, $Y=[16a{B_g}{\chi}^2+(64\pi a{B_g}+3B^2+7Ba+2b)\chi+12\pi(B^2+Ba-2b)]$, $P=24(B-\frac{a}{2})\pi$, $Q=Ba-6a^2+6b$ and $X=(32\pi{B_g}+10B+a)$.

\begin{figure}
\centering
\includegraphics[width=6cm]{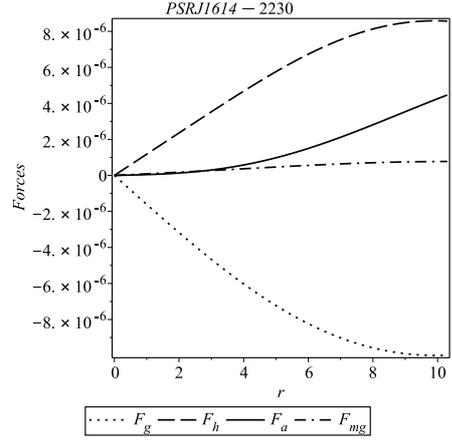}
\caption{Variation of different forces w.r.t. the radial coordinate $r$ for the strange star candidate $PSRJ$~1614-2230.}\label{3.force}
\end{figure}

Fig.~\ref{3.force} shows the variation of different forces with respect to the radius of the star. From the pictorial representation it is
clear that inward pull of $F_g$ is counter balanced by the combined effect of $F_h$, $F_a$ and $F_{mg}$ which acts along outward direction. For $\chi=0$ the usual form of TOV equation as in GR will be retrieved because in this situation the force due to modified gravity vanishes, i.e., effect of coupling vanishes.

\subsubsection{Adiabatic Index}\label{subsubsec5.3.3}
The term adiabatic index $\Gamma$, ratio of two specific heat, represents the stiffness of the EOS for a given density profile~\cite{Hillebrandt1976}. It can be used to study stability of a relativistic and non relativistic fluid sphere. According to Chandrasekhar~\cite{Chandrasekhar1964} the dynamical stability of the stellar model can be checked against an infinitesimal radial adiabatic perturbation. Later on several researchers~\cite{Bardeen1966,Wald1984,Knutsen1988,Herrera1997,Horvat2011,Doneva2012,Mak2013,Silva2015} used the idea to different astrophysical system.

According to this stability condition $\Gamma$ must be greater than $\frac{4}{3}$. For an anisotropic fluid sphere $\Gamma$ can be represented as $\Gamma_{r}$ and $\Gamma_{t}$, the radial adiabatic index and tangential adiabatic index respectively~\cite{Bondi1964}. Chan et al.~\cite{Chan1993} and Heinzmann~\cite{Heintzmann1975} predicted that adiabatic index should exceed $\frac{4}{3}$ inside an anisotropic, relativistic and dynamically stable stellar system. $\Gamma_{r}$ and $\Gamma_{t}$ can be expressed mathematically as follows
\begin{eqnarray}
&\qquad\hspace{+1.1cm}\Gamma_r=\frac{\rho^{eff}+p_r^{eff}}{p_r^{eff}}\left[\frac{dp_r^{eff}}{d\rho^{eff}}\right], \\ \label{eq3.25e}
&\qquad\hspace{+1.1cm}\Gamma_t=\frac{\rho^{eff}+p_t^{eff}}{p_t^{eff}}\left[\frac{dp_t^{eff}}{d\rho^{eff}}\right].  \label{eq3.25f}
\end{eqnarray}

\begin{figure}
\centering
\includegraphics[width=6cm]{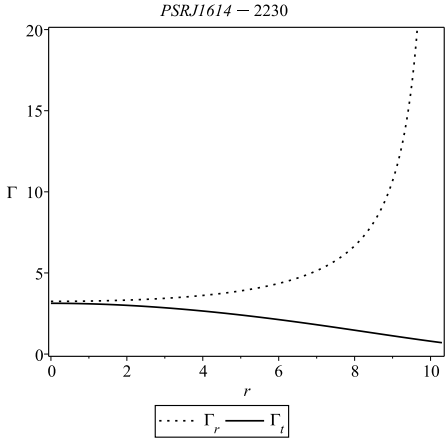}
\caption{Variation of the adiabatic indices $\Gamma_{r}$ and $\Gamma_{t}$ w.r.t. the
radial coordinate $r$ for the strange star candidate $PSRJ$~1614-2230.}\label{3.adiabatic}
\end{figure}

Graphical representations of radial and tangential adiabatic indices are shown in Fig.~\ref{3.adiabatic}. One can note that throughout the
stellar distribution adiabatic indices are greater than $\frac{4}{3}$ and hence the model represents a stable configuration.

\subsubsection{Equation of State (EOS)}\label{subsubsec5.3.4}
The simplest form of the barotropic equation of state can be represented as $p_i=\omega_i\rho$, where $\omega_i$ are the parameters
along the radial and transverse direction. Since we have considered the anisotropic strange star under $f(R,T)$ gravity our radial and transverse EOS parameter will be given by as follows:
\begin{eqnarray} \label{eq3.24}
\omega_r^{eff}=\frac{p_r^{eff}}{\rho^{eff}},~~~ \omega_t^{eff}=\frac{p_t^{eff}}{\rho^{eff}}.
\end{eqnarray}

\begin{figure}
\centering
\includegraphics[width=6cm]{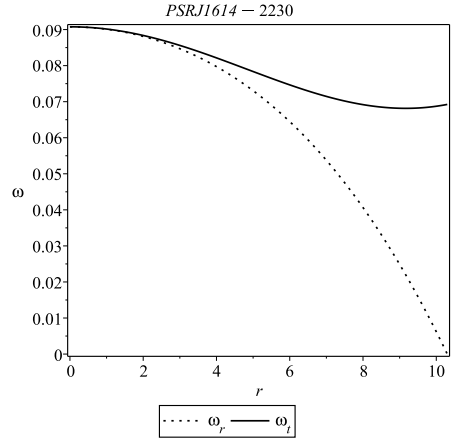}
\caption{Variation of the EOS parameter ${\omega_r^{eff}}$ and ${\omega_t^{eff}}$ w.r.t. the
radial coordinate $r$ for the strange star candidate $PSRJ$~1614-2230.}\label{3.eos.}
\end{figure}

For simplicity we have considered the fluid distribution to be spatially homogeneous, though different literature
survey~\cite{Chervon2000,Zhuravlev2001,Peebles2003,Usmani2008} shows the possibility of space and time dependency of $\omega$.
From the graphical representation of EOS~(Fig. \ref{3.eos.}) it is clear that $0<\omega_i<\frac{1}{3}$, i.e., the matter content of our proposed model is non-exotic in nature~\cite{Shee2016} and decreases monotonically towards the surface of the star.

\subsubsection{Harrison-Zel$'$dovich-Novikov static stability criteria}\label{subsubsec5.3.5}
According to Harrison et al.~\cite{Harrison1965} and Zel$'$dovich and Novikov~\cite{Zeldovich1971} the adiabatic index of a slowly deformed matter is comparable with that of a pulsating star. This result states that for a stable configuration the nature of the mass will be increasing with respect to the central density  (i.e., $\frac{dM}{d\rho_c}>0$) and for an unstable configuration $\frac{dM}{d\rho_c}<0$ which can be provided as follows
\begin{eqnarray}
&\qquad\hspace{-0.5cm}M(\rho_c)=\frac{(\Psi-80\pi\rho_c+15B+36B_g)R^3}{2[\Psi R^2-80\pi R^2\rho_c+15BR^2+36R^2B_g-36\pi-27]}, \\
&\qquad\hspace{-0.5cm}\frac{dM\rho_c}{d\rho_c}=\frac{72(12\pi+5)\pi{R}^3(4\pi+3)}{[\Psi R^2-80\pi{R}^2\rho_c+15BR^2+36R^2B_g-36\pi-27]^2},
\end{eqnarray}
where $\Psi=-36\pi R^2b+192\pi^2B_g-192\pi^2\rho_c-27R^2b+36B\pi+192\pi B_g$.

\begin{figure}
\centering
\includegraphics[width=6cm]{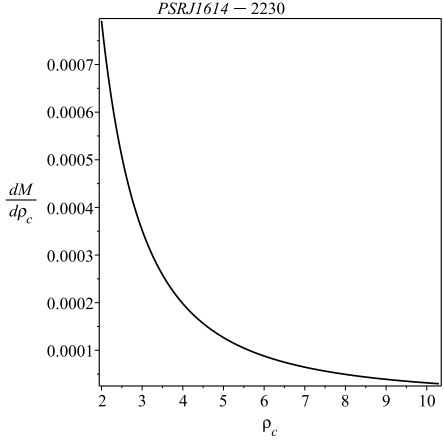}
\caption{Variation of $\frac{dM}{d\rho_c}$ w.r.t. $\rho_c$ for the strange star candidate $PSRJ$~1614-2230.}\label{3.HZN}
\end{figure}

Fig.~\ref{3.HZN} conveys that though $\frac{dM}{d\rho_c}$ decreases with the increase of $\rho_c$, it remains always positive throughout the stellar structure. So our model is stable according to Harrison-Zel$'$dovich-Novikov condition.

\subsection{Compactification factor and surface redshift}\label{sec6}
We can calculate the effective mass for our static, spherically symmetric and anisotropic fluid distribution through the relation
\begin{eqnarray}
M^{eff} &=& \int _{0}^{\Re}4\pi r^2\rho^{eff} dr \nonumber\\
&=&\int _{0}^{\Re}4\pi r^2\left[\rho+\frac{\chi}{4\pi}\left(2\rho-\frac{p_r+2p_t}{3}\right)\right] dr \nonumber\\
&=&m+\int _{0}^{\Re}r^2\chi\left(2\rho-\frac{p_r+2p_t}{3}\right)dr. \label{eq3.30}
\end{eqnarray}

\begin{figure}[!htp]
\centering
\includegraphics[width=6cm]{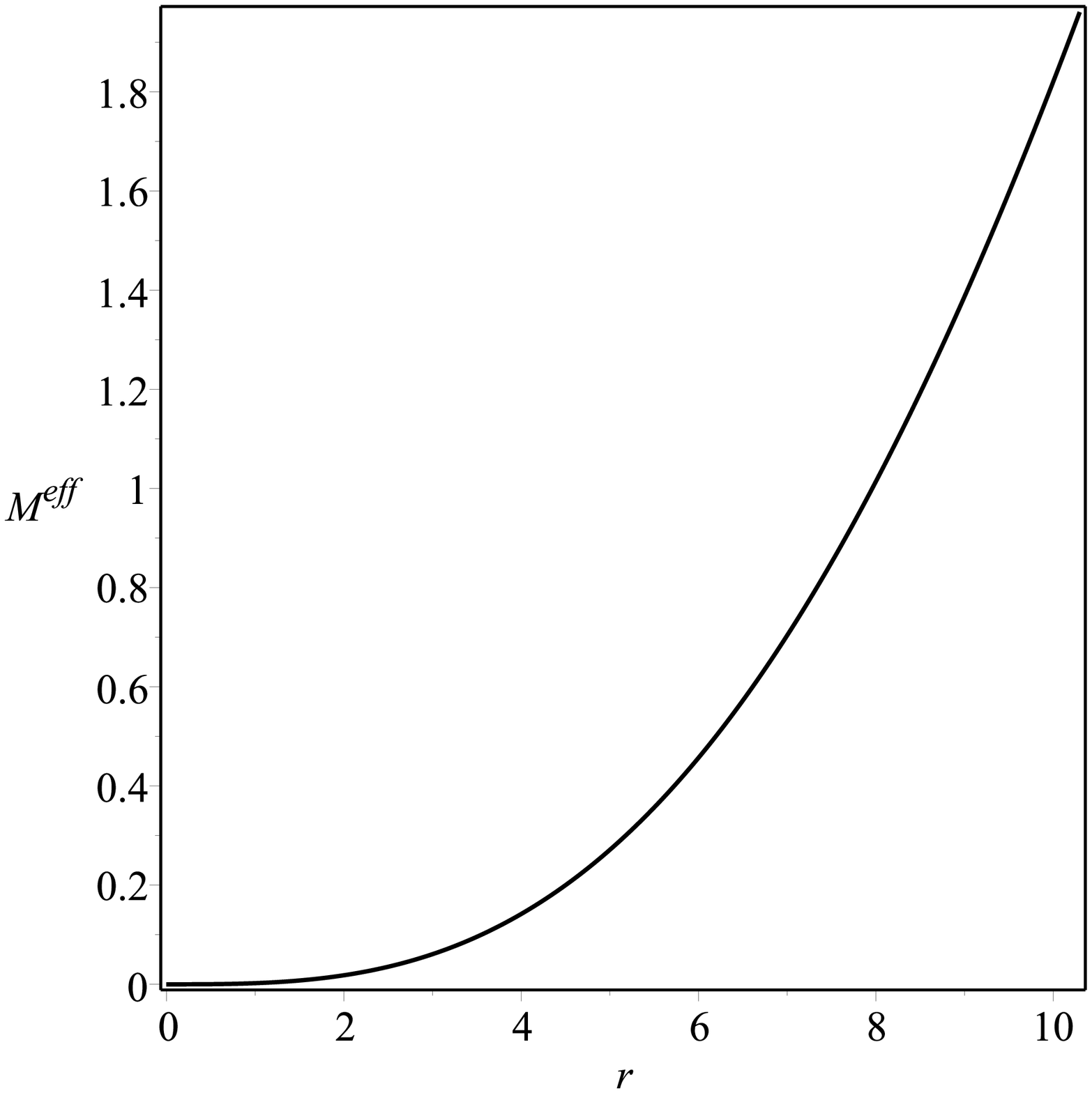}
\includegraphics[width=6cm]{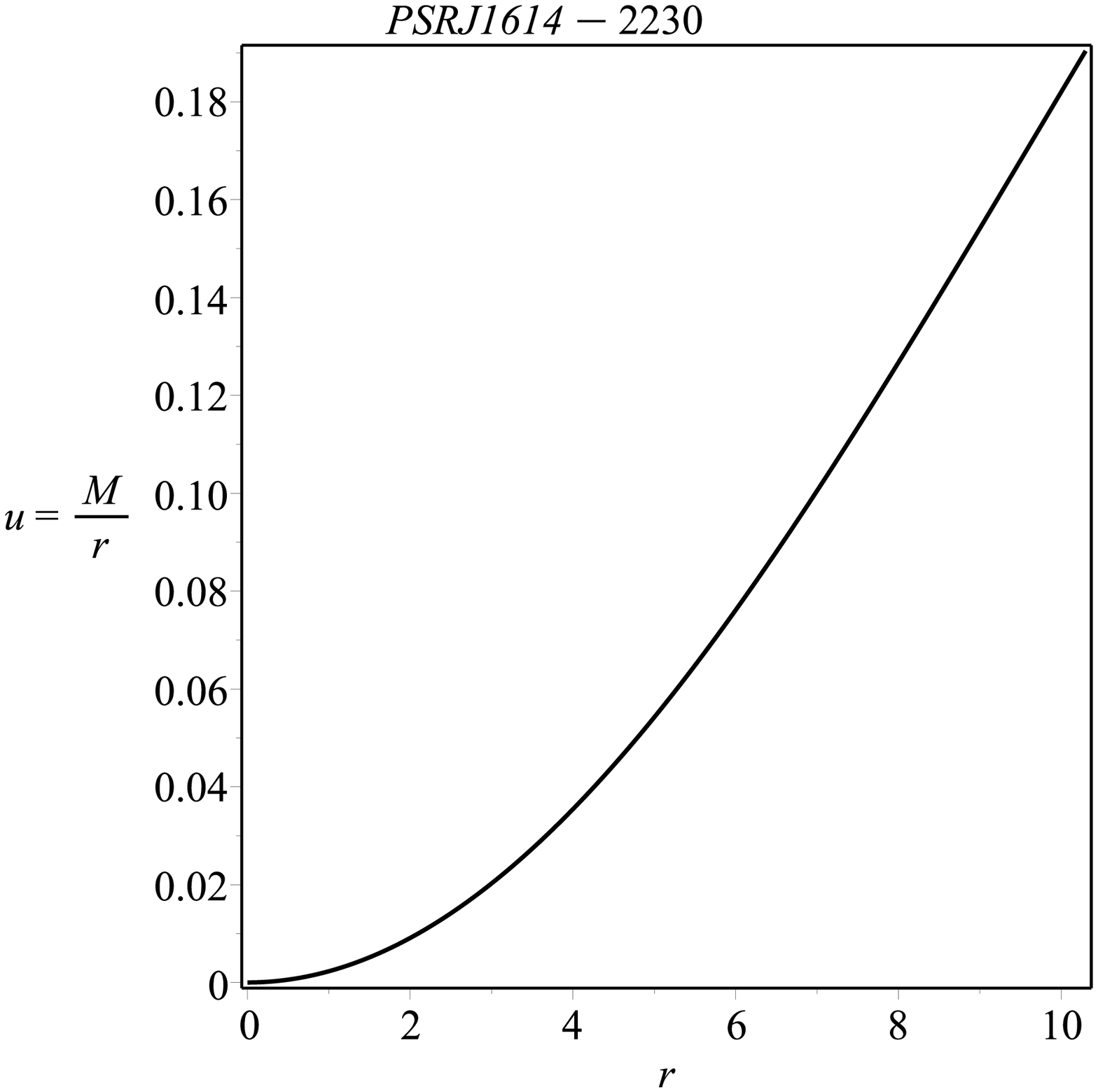}
\caption{Variation of mass (upper panel), compactness (lower panel) w.r.t. the radial coordinate $r$ for the strange star candidate $PSRJ$~1614-2230.}\label{3.mass}
\end{figure}

From the Fig.~\ref{3.mass} (left panel) we can see that  $M^{eff}$ increases monotonically with the radius of the star and as
$r \rightarrow 0$, $M^{eff} \rightarrow 0$. We can observe in Eq.~(\ref{eq3.30}) that effect of $\chi$ is very much prominent. The term with $\chi$ represents mass due to modified gravity, without this term we will get the traditional mass as in the classical physics. As $\chi \rightarrow 0$ the mass due to modified gravity will also be zero.

The factor $\frac{M}{\Re}$ defines the compactification of the stellar system. Buchdahl~\cite{Buchdahl1959} in one of his pioneering work derived an upper limit for the allowed mass to radius ratio, i.e., $\frac{M}{\Re}<\frac{4}{9}$, for a static spherically symmetric perfect fluid star. Later on Mak et al.~\cite{Mak2001} generalized it for charged sphere. Jotania and Tikekar~\cite{Jotania2006} classifies the stellar objects into different categories on the basis of the values of $\frac{M}{\Re}$ as follows - (i) normal star: $M/\Re \sim 10^{-5}$, (ii) white dwarf: $M/\Re \sim 10^{-3}$, (iii) neutron star: $10^{-1}<M/\Re <1/4$, (iv) ultra dense compact star: $1/4<M/\Re<1/2$ and (v) black hole: $M/\Re=1/2$.

The expression of the compactification factor is given by
\begin{eqnarray}
&\qquad\hspace{-0.8cm}u(r)=\frac{M^{eff}(r)}{\Re}=\frac{1}{\Re}[m+\int _{0}^{\Re}r^2\chi\left(2\rho-\frac{p_r+2p_t}{3}\right)dr].  \label{eq3.31}
\end{eqnarray}

The graphical representation of the above factor for our model is shown in Fig.~\ref{3.mass} (lower panel). We can note very clearly that it is a monotonic increasing  function of the radius of the star and its maximum value implies that our model corresponds to an ultra dense compact object. Here also the effect of $\chi$ is prominent. It plays a significant role in determining the compactification factor of the strange stars in $f(R,T)$ gravity.

The surface redshift ($Z_s$) and gravitational redshift ($Z$) are represented respectively as
\begin{eqnarray}
Z_{{s}}&=&{\frac {1}{\sqrt {1-2\,u}}}-1, \\ \label{eq3.32}
Z&=&e^{\frac{-\nu(r)}{2}}-1={\frac {1}{\sqrt {{{\rm e}^{B{r}^{2}+2\,\ln  \left( C \right) }}}}}-1. \label{eq3.32}
\end{eqnarray}

Barraco and Hamity~\cite{Barraco2002} proved that for an isotropic star and in absence of cosmological constant $Z_{s}<2$. According to
Bohmer and Harko~\cite{Bohmer2006} surface redshift for an anisotropic star can reach maximum higher value $Z_{s} \leq 5$, in presence
of cosmological constant. Though according to Ivanov~\cite{Ivanov2002} maximum acceptable value of surface redshift will be $5.211$.

\begin{figure}[!htp]
\centering
\includegraphics[width=6cm]{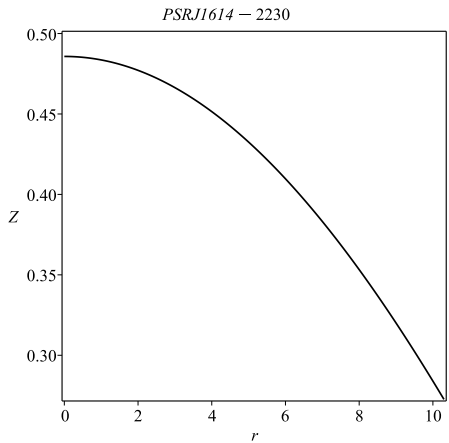}
\caption{Variation of gravitational redshift w.r.t. the radial coordinate $r$ for the strange star candidate $PSRJ$~1614-2230.}\label{3.redshft}
\end{figure}

We have plotted the gravitational redshift $Z$ for our model in Fig.~\ref{3.redshft}, where it is clear that it decreases with the radius of the star. It's value on the surface of the star, i.e., the surface redshift is $0.271$ which strongly confirms the acceptance of our model as a  strange star.

\section{Discussions and concluding remarks}\label{sec7}

\begin{table*}[!htp]
\setlength\tabcolsep{12pt} \centering \caption{Mass and radius of different strange star candidates~\cite{Rahaman2014} }
      \begin{tabular}{cccccccc}
\hline        Case& Stars               & Mass             & Radius     & $\frac{M}{\Re}$  & $B_g (\chi=1)$          & $B_g (\chi=0)$          & $Z_s$  \\
                  &                     &$(M_\odot)$       &($\Re$ in km) &                &{\small{(MeV/{fm}$^3$)}}   &{\small{(MeV/{fm}$^3$)}}   &        \\
\hline        I   & $PSR~J~1614~2230$   & 1.97$\pm$0.04    & 10.3       & 0.191          & 47.55                   & 68.01                   & 0.271  \\
              II  & $Vela~X-1$          & 1.77$\pm$0.08     & 9.99      & 0.177          & 47.64                   & 68.23                   & 0.2432 \\
              III & $PSR~J~1903+327$    & 1.667$\pm$0.021   & 9.82      & 0.17           & 47.67                   & 68.4                    & 0.23   \\
              IV  & $Cen~X-3$           & 1.49$\pm$0.08     & 9.51      & 0.157          & 47.59                   & 68.41                   & 0.206   \\
              V   & $SMC~X-1$           & 1.29$\pm$0.05     & 9.13      & 0.1413         & 47.334                  & 68.17                   & 0.18 \\
\hline \label{Table1}
\end{tabular}
\end{table*}

\begin{table*}[!htp]
\setlength\tabcolsep{12pt} \centering \caption{Determination of model parameters $a$, $b$, $C$, $B$, central density, surface density, radial pressure for different strange star candidates}
\begin{tabular}{cccccccc}
\hline
         Case  & $a$       & $b$        & $B$            &$C$                      & $\rho^{eff}(r=0)$       & $\rho^{eff}(r=\Re)$       & $p_r^{eff}(r=0)$ \\
               &{\small{(km$^{-2}$)}}&{({km}$^{-4}$)} &{\small{({km}$^{-2}$)}}&   &{\small{({gm/cm}$^3$)}}   &{\small{({gm/cm}$^3$)}}  &{\small{({dyne/cm}$^2$)}}  \\
\hline   I     & 0.00459   & 0.0000118  & 0.00292        & 0.6731                  & 7.412$\times10^{14}$    & 4.85$\times 10^{14}$    & 6.0331$\times 10^{34}$ \\
         II    & 0.0044    & 0.000011   & 0.00275        & 0.7005                  & 7.1131$\times10^{14}$   & 4.871$\times 10^{14}$   & 5.286$\times 10^{34}$  \\
         III   & 0.0043    & 0.0000106  & 0.00267        & 0.715                   & 6.967$\times10^{14}$    & 4.879$\times 10^{14}$   & 4.924$\times 10^{34}$ \\
         IV    & 0.0042    & 0.0000099  & 0.002523       & 0.7393                  & 6.71$\times10^{14}$     & 4.88$\times 10^{14}$    & 4.311$\times 10^{34}$ \\
         V     & 0.00397   & 0.0000091  & 0.002363       & 0.7676                  & 6.41$\times10^{14}$     & 4.863$\times 10^{14}$   & 3.653$\times 10^{34}$  \\
\hline \label{Table2}
\end{tabular}
\end{table*}

Motivation of this work is to study the relativistic strange star under the framework of $f(R,T)$ gravity where spherically symmetric
spacetime is of Tolman-Kuchowicz type. With the help of SQM distribution, demonstrated as in the famous MIT bag EOS~[Eq.~(\ref{eq3.12})], we explore strange quark star. As anisotropic matter distribution is considered here, incorporation of the EOS does not make the system over determined. Solving EFEs [Eqs.~(\ref{eq3.9})-(\ref{eq3.11})] in modified gravity, using Tolman-Kuchowicz metric as well as MIT bag EOS, we have obtained different physical parameters. Numerical values of different physical quantities have been calculated and shown in tabular form and their graphical variations have been shown w.r.t. the radial co-ordinate $r$. Though graphical representation has been made for the candidate $PSR~J~1614~2230$, we have shown the respective values for different strange star candidates, like $Vela~X-1$, $PSR~J~1903+327$, $Cen~X-3$ and $SMC~X-1$ also.

Geometry of the space-time can be specified by the metric potentials $\nu(r)$ and $\lambda(r)$. From variation of $\nu(r)$ and $\lambda(r)$
in Fig.~\ref{3.pot}, it's clear that $e^{\nu(r)}(r=0)\neq0$ and positive Also $e^{\lambda(r)}(r=0)=1$, which are the necessary conditions to
achieve the non-singular solutions from physical as well as geometrical aspects. Both the functions $e^{\nu(r)})$ and $e^{\lambda(r)}$
increases non-linearly toward the surface from the center.

In $f(R,T)$ gravity, a constant $\chi$ arises due to the coupling between matter and geometry. Assuming $\chi=1$ in anisotropic condition,
we can calculate the bag constant $B_g$ which remains in the range (47-48)~MeV/fm$^{3}$, though the predicted range for stable SQM matter distribution is (55-75)~MeV/fm$^{3}$~~\cite{Farhi1984,Alcock1986} in GR. However, very recently Aziz et. al~\cite{Aziz2019} have discussed about the possible wide range (41.58-319.31)~MeV/fm$^{3}$ for $B_g$ on the basis of observational data for different strange star candidates. In our model the coupling parameter $\chi$ effectively reduces to $B_g$ in case of $f(R,T)$ gravity. Incorporating $\chi=0$ in anisotropic condition, we get $B_g$ in the range (68-69)~MeV/fm$^{3}$. Biswas et al.~\cite{Biswas2018} have previously shown that $\chi$ is blameworthy behind this effective reduction of $B_g$. The coupling constant plays a very crucial role to study the strange star. Depending on its value we can study different aspect of Bag constant.

Bauswein and his co-workers~\cite{Bauswein2009,Bauswein2010} predicted a strange star-strange star merger process which increases the probability of SQM hypothesis if one can detect Gravitational Wave from this merging phenomenon. It is to note that some simple features of the GW signals may reveals whether Strange Star (SS) or Neutron Star (NS) merger had produced the emission. In case of SS mergering, the maximal frequency during the inspiral and the frequency of ringdown of the postmerging remnant are higher than those of NS merging. A particular choice of the EOS will make the frequencies similar for SS and NS merging procedure. In such a case one can take the help of Gravitational Wave luminosity spectrum to reveal the features. As a result to differentiate between NS and SS merging one can use the ratio of energy emitted during the inspiral phase to the energy radiated away in the postmerger stage. Hopefully, the upcoming advanced GW detectors LIGO~\cite{Abbott2016a,Abbott2016b} and VIRGO~\cite{Acernese2006} may provide more valuable data regarding the signals from high density compact object binaries~\cite{Kalogera2004}.

Other forms of self-bound matter like pion condensation may lead to the stellar objects similar to SS~\cite{Voskresenski1977,Migdal1990,Haensel2007}. So we can generalize our study to these states of matter also provided we have the knowledge of the specific EOS. In addition to GW measurements the cooling history of NS, SS and self-bound pion condensed stars are different~\cite{Weber1999,Blaschke2000,Yakovlev2004,Grigorian2005}. Since merger remnants are the promising source to develop better understanding regarding the origin of GW spectra, one can claim in future a better understanding of SS and BH through SS-BH merger.

Now the current study can be summarized as follows:

\textbf{(i) Density and pressure:} The effective density, effective radial and tangential pressure are maximum at the center of the star and
decrease monotonically towards the surface of the star, shown in the Fig.~\ref{3.pressue} as well as in Table~2.  However, it is to be noted that the effective density
at the surface of the star reduces to $34.56~\%$ of its value at the center. The anisotropic stress is minimum at the center of the star and
increases non-linearly with the radius of the star and acquire the maximum value at the surface of the system. The tremendous pressure at the center make the system ultra compact object. The spheroidicity in our system arises due to this maximum anisotropic force at the surface of the star. The pictorial variation of EOS Fig.~\ref{3.eos.} states clearly that  $0<\omega_i<\frac{1}{3}$, i.e., our system consists of non-exotic matter.

\textbf{(ii) TOV equation:} In $f(R,T)$ gravity, our proposed model satisfies the force condition, i.e., the generalized TOV equation [Eq.~\ref{eq3.25}]. From Fig.~\ref{3.force}, it is distinct that the effect of gravitational force ($F_g$) stabilizes the unified effect of hydrostatic
force ($F_h$), anisotropic force ($F_a$) and the newly modified gravity force ($F_{mg}$), originated due to the matter-geometry coupling in $f(R,T)$ gravity.

\textbf{(iii) Energy Conditions:} Our system is consistence with all the energy conditions, e.g. NEC, WEC, SEC and DEC for both the radial and tangential cases of pressure. Variations of all these conditions have been shown graphically (Fig.~\ref{3.energy}) which re-confirms the physical stability and acceptability of proposed model.

\textbf{(iv) Stability:} To examine the stability of the system, we have checked Herrera's cracking condition and the causality conditions as
$\vert v^2_{rs}-v^2_{ts} \vert \leq 1$, $0<v^2_{rs}<1$ and $0<v^2_{ts}<1$. Fig.~\ref{3.Hererra} also manifest that all the inequality criteria are well satisfied This enhances the physical acceptability of the system with the background of sound velocity of the system.

From Fig.~\ref{3.adiabatic}, it's clear that both the adiabatic indices (i.e., $\Gamma_{r}$ and $\Gamma_{t}$) remain within the critical value $\frac{4}{3}$~\cite{Bondi1964} through out the system which establish that infinitesimal adiabatic perturbation can't distort the stability of the model. Harrison-Zel$'$dovich-Novikov's static stability criteria is well satisfied, Fig.~\ref{3.HZN} shows that $\frac{dm}{d\rho_c}$ is always positive throughout the system.

\textbf{(v) Buchdahl condition:} Variation of the effective mass has been featured in Fig.~\ref{3.mass} (upper panel) which emphasizes the regularity of $M(r)^{eff}$ as for $r\rightarrow0$, $M(r)^{eff}\rightarrow0$. According to Buchdahl condition~\cite{Buchdahl1959}, for static spherically symmetric perfect fluid distribution the mass radius ratio, i.e., $\frac{M}{\Re}\leq \frac{4}{9}$ (Table~\ref{Table1}) which is also satisfied for every strange star candidate.

\textbf{(vi) Compactification factor and redshift:} We have studied the compactification factor $u$ and shown graphically in Fig.~\ref{3.mass} (lower panel), where the revealed features highly recommend for a strange star candidate. The gravitational redshift (Fig.\ref{3.redshft}) reduces continuously from center to surface. The surface redshift is high enough ($\sim 0.18-0.27$) for all the stars which strongly indicates the possibility of stable star configuration.

At this point we would like to conduct a comparative study between our work in $f(R,T)$ gravity and that of the investigations by other authors~\cite{Cruz-Dombriz2006,Astashenok2015,Capozziello2016}, especially by Asthashenok et al.~\cite{Astashenok2015} in $f(R)$ gravity which are as follows:\\
(i) In $f(R)$ gravity, $f(R)$ contains 1st and higher order terms of scalar curvature $R$, connected through a constant parameter $\alpha$. Setting $\alpha=0$ gives the simpler form as $f(R)=R$. But in our work of $f(R,T)$ gravity, where Ricci scalar $R$ and the trace of the energy-momentum tensor $T$ are connected through a coupling constant parameter $\chi$, arises due to the matter-geometry coupling.

(ii) In the above study, $\alpha$ is the key factor and different results with the variation of $\alpha$ have been investigated in their study whereas in our case the key factor is $\chi$ and setting $\chi=0$ gives the results for GR as well as non-zero $\chi$ represents the alternative gravity.

(iii) The study~\cite{Astashenok2015} is based on general relativistic stars like neutron stars and quark stars. It's not explicitly clear whether the model is energetically stable or not. Our study is specifically based on strange quark stars where we have calculated different unknown parameters like the matter density, radial and tangential pressure, value of EOS parameter, redshift etc. Besides, consistency of our model with various stability conditions at a time, is also worthy to mention.

(iv) Though Asthashenok et al.~\cite{Astashenok2015} in their study have also mentioned about the MIT bag model EOS and and hence bag constant $B_g$. But, here we have calculated the range for $B_g$ in $f(R,T)$ gravity. We have also shown that value of $B_g$ reduces in $f(R,T)$ due to the effect of $\chi$. Setting $\chi=0$ one can get the higher value of $B_g$ which exactly matches to the predicted range~\cite{Farhi1984,Alcock1986}.  
 
In connection to the above comparative discussion we would also like to make comments on the current article regarding strange stars that the study shows remarkably different results in comparison to the earlier works~\cite{Rahaman2012,Bhar2015,Deb2017,Deb2018a,Deb2018b}. Here, one can find a very interesting feature that the value of bag constant $B_g$ gets effected under modified gravity. However, the new results can be mentioned as follows:

(i) In this study, we have shown the existence of stable strange quark stars with lower bag value in the range (47-48) Mev/fm$^3$ whereas earlier works obtained~\cite{Rahaman2012,Bhar2015} or assumed~\cite{Deb2017,Deb2018a,Deb2018b} $B_g$ in the higher range for strange star models in GR~\cite{Rahaman2012,Bhar2015} as well as in the modified theories of gravity~\cite{Deb2017,Deb2018a,Deb2018b}. Our calculations in this study, therefore, clearly illustrate the effect of $f(R;T)$ gravity. Here, the matter-geometry coupling constant $\chi$, effectively reduces the value of $B_g$. Besides, minimization of the matter-geometry coupling (i.e. $\chi=0$) in this study, gives back the higher values of $B_g$ in the range (55-75) Mev/fm$^3$ as proposed by Farhi~\cite{Farhi1984} and Alcock~\cite{Alcock1986} for stable strange star model.

(ii) Following the literature~\cite{Rahaman2012,Bhar2015,Deb2017,Deb2018a,Deb2018b}, we have studied all the energy conditions, TOV equation, Herrera's cracking conditions, Buchdahl limit, adiabatic index and the EOS parameter. We have also shown the variations of all these
parameters w.r.t. $r$ for strange star candidate PSR J 1614 2230. Our proposed model of anisotropic strange star with Tolman-Kuchowicz metric under modified $f(R,T)$ gravity, shows it's consistency with all the stability criteria and indicates the physical acceptability and stability. Proposed model also satisfy the Harrison-Zel'dovich-Novikov criteria for stability. It's mention worthy that no earlier work~\cite{Rahaman2012,Bhar2015,Deb2017,Deb2018a,Deb2018b}  shows it's consistency with all the static stability criteria as our study does effectively.

We finally comment that our model is completely free from both physical and geometrical singularities and represents a strange quark star with highly dense SQM distribution. It can be observed that under $f(R,T)$ gravity, study of anisotropic strange star using Tolman-Kuchowicz metric, provides such a model which satisfy all the stability criteria and perfect for studying different features of anisotropic strange stars. Along with this, another fascinating feature is the reduction in bag value. The coupling constant $\chi$ caused due to the coupling between matter and geometry, effectively reduces $B_g$ and also the value of $v^2_{rs}$, which is constant $(\frac{1}{3})$ in GR. For both the cases, $\chi=0$ retrieves the result that exactly matches to GR results. However, as our model works well within the low range (47-48)~MeV/fm$^{3}$ for $\chi=1$ and (68-69)~MeV/fm$^{3}$ for $\chi=0$, therefore the present investigation may be generalized by considering more realistic EOS obtained from QCD simulations~\cite{Alford2013} to make it also suitable for the higher range (41.58-319.31)~MeV/fm$^{3}$~\cite{Aziz2019}.

\section*{Acknowledgement}
SR is thankful to the Inter University Centre for Astronomy and Astrophysics (IUCAA) for providing
Visiting Associateship. SB is thankful to DST-INSPIRE [IF~160526] for financial support and all type of facilities
for continuing research work.

\end{document}